\documentclass[12pt,a4paper]{iopart}
\usepackage{iopams}
\usepackage{amssymb}
\usepackage{harvard}
\usepackage{graphicx}
\usepackage{subfigure}
\usepackage{epstopdf}
\usepackage{rotating,booktabs}
\usepackage{multirow}
\usepackage[colorinlistoftodos]{todonotes}
\usepackage{hyperref}

\hypersetup{
breaklinks=true,
pdftitle={Fast synthetic CT generation using a generative adversarial network for general pelvis MR-only radiotherapy},    	
pdfauthor={Matteo Maspero},      
pdfsubject={MR-only treatment planning for general pelvis},   
pdfkeywords={magnetic resonance imaging, cancer, pseudo CT, dose calculations, neuronal network, generative adversarial network, medical imaging},
pdfcreator={mmaspero},   
pdfproducer={UMC Utrecht}}

\begin{document}

\title[Fast sCT generation using a GAN for general pelvis MR-only radiotherapy]{Dose evaluation of fast synthetic-CT generation using a generative adversarial network for general pelvis MR-only radiotherapy}

\author{Matteo Maspero$^{*,1,2,3}$,Mark H F Savenije$^{*,1,2}$, Anna M Dinkla$^{1,2}$, Peter R Seevinck$^{2,3}$, Martijn P W Intven$^{1}$, Ina M Jurgenliemk-Schulz$^{1}$, 
Linda G W Kerkmeijer$^{1}$, Cornelis A T van den Berg$^{1,2}$}

\address{$^*$ The authors equally contributed.}
\address{$^1$ Department of Radiotherapy, University Medical Center Utrecht, Utrecht, the Netherlands}
\address{$^2$ Center for Image Sciences, University Medical Center Utrecht, Utrecht, The Netherlands}
\address{$^3$ Image Science Institute, University Medical Center Utrecht, Utrecht, The Netherlands}
\ead{m.maspero@umcutrecht.nl, matteo.maspero.it@gmail.com}

\begin{indented}
\item[]\today
\end{indented}


\begin{abstract} 
To enable magnetic resonance (MR)-only radiotherapy and facilitate modelling of radiation attenuation in humans, synthetic CT (sCT) images need to be generated.
Considering the application of MR-guided radiotherapy and online adaptive replanning, sCT generation should occur within minutes.
This work aims at assessing whether an existing deep learning network can rapidly generate sCT images to be used for accurate MR-based dose calculations in the entire pelvis.

A study was conducted on data of 91 patients with prostate (59), rectal (18) and cervical (14) cancer who underwent external beam radiotherapy acquiring both CT and MRI for patients' simulation.
Dixon reconstructed water, fat and in-phase images obtained from a conventional dual gradient-recalled echo sequence were used to generate sCT images.
A conditional generative adversarial network (cGAN) was trained in a paired fashion on 2D transverse slices of 32 prostate cancer patients.
The trained network was tested on the remaining patients to generate sCT images.
For 30 patients in the test set, dose recalculations of the clinical plan were performed on sCT images.
Dose distributions were evaluated comparing voxel-based dose differences, gamma and dose-volume histogram (DVH) analysis.

The sCT generation required 5.6~s and 21~s for a single patient volume on a GPU and CPU, respectively.
On average, sCT images resulted in a higher dose to the target of maximum 0.3\%.
The average gamma pass rates using the 3\%,3mm and 2\%,2mm criteria were above 97 and 91\%, respectively, for all volumes of interests considered. All DVH points calculated on sCT differed less than $\pm$2.5\% from the corresponding points on CT.

Results suggest that accurate MR-based dose calculation using sCT images generated with a cGAN trained on prostate cancer patients is feasible for the entire pelvis.
The sCT generation was sufficiently fast to be integrated into an MR-guided radiotherapy workflow.

\end{abstract}

\submitto{Phys. Med. Biol.}

\section{Introduction}

``Magnetic resonance (MR)-only'' radiotherapy refers to a radiotherapy workflow in which patient simulation and dose calculation are performed using only MR images.
This workflow has been proposed to exploit the soft tissue contrast offered by magnetic resonance imaging (MRI) without recurring to inter-modality
registration and thus reducing possible systematic errors in target definition \cite{Nyholm2009}.
Also, MR-only radiotherapy offers practical and logistical advantages reducing the overall treatment cost \cite{Devic2012}, workload \cite{Karlsson2009}
and patient exposure to ionising radiation \cite{Schmidt2015}.
In addition, MR-only is of interest considering the advent of MR-guided radiotherapy (MRgRT) systems \cite{Lagendijk2014,Low2017},
where it may be exploited to perform online daily replanning based on the anatomy acquired with MRI before irradiation.

However, dose calculation cannot be performed directly on MR images since no correlation has been demonstrated between the nuclear magnetic properties, on which MRI depends,
and the electron density, which is the property used to model radiation attenuation in humans \cite{Brown2014}.
Several groups have proposed methods to automatically generate the so-called
``synthetic'' CT (sCT)\footnote{In the literature, the sCT images are also called ``pseudo-CT'' or ``substitute CT''.}
images to enable accurate MR-based dose calculation \cite{Owrangi2018}.

When focusing on the pelvis area, many methods have been proposed and evaluated for radiotherapy of prostate cancer \cite{Edmund2017,Johnstone2017}, showing dose deviations below 2\% with respect to CT-based dose calculations.
Only three contributions investigated the accuracy of MR-based dose calculations for locations other than prostate within the pelvic area
\cite{Kemppainen2017,Liu2017,Wang2018}.
Moreover, no attention has been specifically dedicated to the time required to generate sCT, which should be of the order of minutes
to allow daily replanning during MRgRT as, for example, underlined by \citeasnoun{Raaymakers2017}.

Recently, deep learning-based sCT generation has been presented \cite{Nie2017,Han2017,Wolterink2017}
enabling sCT generations of a full 3D volume in a minute. 
\citeasnoun{Han2017} employed, for the first time, deep learning to generate sCT images using only a generative model, e.g. a U-net. 
Generative adversarial networks (GANs) have been used to synthesise medical images after training in a paired and an unpaired fashion by \citeasnoun{Nie2017} and \citeasnoun{Wolterink2017}, respectively.
They both obtained promising results, especially compared to networks in which only the generator was adopted \cite{Han2017}.
However, in none of these contributions, the sCT images were evaluated for dosimetric accuracy, which is the relevant metric
for radiotherapy purposes.
In this sense, we presented a dosimetric evaluation on deep learning-based sCT images for brain patient using a dilated convolutional network \cite{Dinkla2018}.
So far, no study has reported evaluation of dose calculation accuracy for deep-learning based sCT in the pelvis.

By interpreting the generation of sCT images as an image-to-image problem,
we aim at assessing whether an existing deep learning network
can generate sCT images that enable accurate MR-based dose calculation using a conventional MR sequence in the pelvic area.
We used a conditional generative adversarial network (cGAN) motivated by the results obtained by \citeasnoun{Isola2016} who showed how this 
approach can promptly solve numerous image-to-image translation problems \cite{Litjens2017}. 
More specifically, GANs are networks constituted by a generator and a discriminator network.
Generator and discriminator are jointly trained, particularly aiming at generating
realistically looking images by exploiting the capability of the 
discriminator network to discern between real and fake images \cite{Goodfellow2014,Isola2016}.
%

In this work, we trained an existing GAN \cite{Isola2016} with paired MRI-CT data to learn the generation of sCT images using multi-contrast Dixon reconstructed
MR images from conventional multi-echo gradient echo.
Training was performed on prostate cancer patients' images only.
Finally, sCT images were evaluated for MR-based dose calculations for patients
with prostate, rectal and cervical cancer.

\section{Materials and methods}

\subsection{Patient data collection}
\label{sec:patcoll} 

This study was conducted on a total of 91 patients: 59 prostate, 18 rectal and 14 cervical cancer patients
who had no hip implants and underwent external beam radiotherapy. Patients simulation was performed using both on CT and MRI images acquired between March 2016 and April 2017.

Fifty-nine patients were diagnosed with low, intermediate, high-risk prostate carcinoma stage T1c-T3b. 
Their mean age was 69.6$\pm$5.1 years ($\pm 1\sigma$; range 59.8-82.9).
Three intra-prostatic cylindrical gold fiducial markers were inserted in these patients for position verification purposes.
Prostate cancer patients underwent 5-beam 10 MV intensity-modulated radiotherapy (IMRT) with a prescribed dose of 35x2.2~Gy to prostate and macroscopic tumour and 35x2.0~Gy to seminal vesicles.

Eighteen patients, of whom 5/18=27.8\% female, were diagnosed with intermediate and high-risk rectal cancer staged T2-T4.
These patients were treated for neoadjuvant therapy with three fractionating regimes: short course treatment delivering 5x5 Gy (2), and long-course treatment 25x2.0 Gy
without (14) and with (2) an integrated boost on extramesorectal pathological nodes of 25x2.4 Gy.
All patients in this group were irradiated with volumetric modulated arc therapy (VMAT)
consisting of two coplanar arcs of 10 MV between 50​$^{\circ}$ and 310​$^{\circ}$.

Fourteen patients were diagnosed with low, intermediate, high-risk cervical cancer staged T1-T4.
Their mean age was 51.4$\pm$15.1 years (range 29.1-83.0 years).
These patients underwent external beam radiotherapy with 10 MV VMAT with a 360$^{\circ}$ irradiation
arc and the following dose schemes: 25x1.8~Gy (2), 25x2.2~Gy with an integrated boost in the pelvic pathological nodes (3) and 25x2.3~Gy with
an integrated boost in the common iliac and para-aortic region (9).

For all patients with prostate and rectal cancer, 3T MRI (Ingenia MR-RT, v 5.7.1, Philips Healthcare, The Netherlands) was acquired within 2.5 hours
the CT (Brilliance Big Bore, Philips Healthcare, Ohio, USA).
For the cervical cancer patients, time between imaging protocols was up to one week.  
All patients were asked to drink between 200 and 300 ml of water one hour before the scans and after emptying the bladder (and rectum in the case of prostate cancer patients).
Patients were positioned using a flat table and knee wedges.
CT scans were performed with the following imaging parameters: 120~kV, 923~ms exposure time, 121-183~mA tube current,
512x512 in-plane matrix, and 3~mm slice thickness. In-plane resolution was variable depending on the field of view (FOV) used, with a typical pixel size of 1x1~mm$^2$ and
maximum size of 1.2x1.2~mm$^2$. In the inferior-superior direction, the size of the FOV was variable ranging 33-77~cm. 

To simulate treatment positioning, patients were marked with at least three skin tattoos at the CT scanner, which were then used to reposition the patient at the MR scanner
with the aid of a laser system (Dorado3, LAP GmbH Laser Applikationen, Germany).
MRI was acquired using anterior and posterior phased array coils (dS Torso and Posterior coils, 28 channels, Philips Healthcare, The Netherlands).
To avoid compression of the patients, two in-house-built bridges supported the anterior coil.

For the generation of MR-based sCT images, a dual echo three-dimensional (3D) cartesian radio-frequency
spoiled gradient-recalled echo sequence was acquired with the following parameters: 1.2/2.5~ms echo times,
3.9~ms repetition time, 10$^{\circ}$ flip angle, 552x552x300~mm$^3$ FOV\footnote{Expressed in terms of anterior-posterior, right-left and superior-inferior directions.},
anterior-posterior as the readout direction (frequency
encoding), 284x281x120 acquisition matrix, 1.05x1.05x2.5~mm$^3$ reconstructed voxel,
1083~Hz/px bandwidth and 2 min 13 s acquisition time.
A Dixon reconstruction \cite{Dixon1984,Eggers2011} was performed obtaining in-phase, fat, and water images.
This sequence was originally acquired to generate sCT for sole prostate patients with a proprietary method called MR for calculating attenuation (MRCAT, Philips Healthcare, The Netherlands)
as presented in \citeasnoun{Tyagi2016} and \citeasnoun{Maspero2017Quant}. In this work, MRCAT was used to automatically identify air regions based on the in-phase, fat and water images to avoid laborious manual segmentation during preparation of the training data.
Identification of air regions was performed as specified in the following section
to ensure consistency
of air locations between CT and MR images during the training of the network.

Delineations of the target volumes and organs at risk (OARs) were performed by radiation oncologists. 

\subsection{The network}
\label{sec:sCTimageproc}
\begin{figure}[!t]
\begin{center}
\includegraphics[width=14cm]{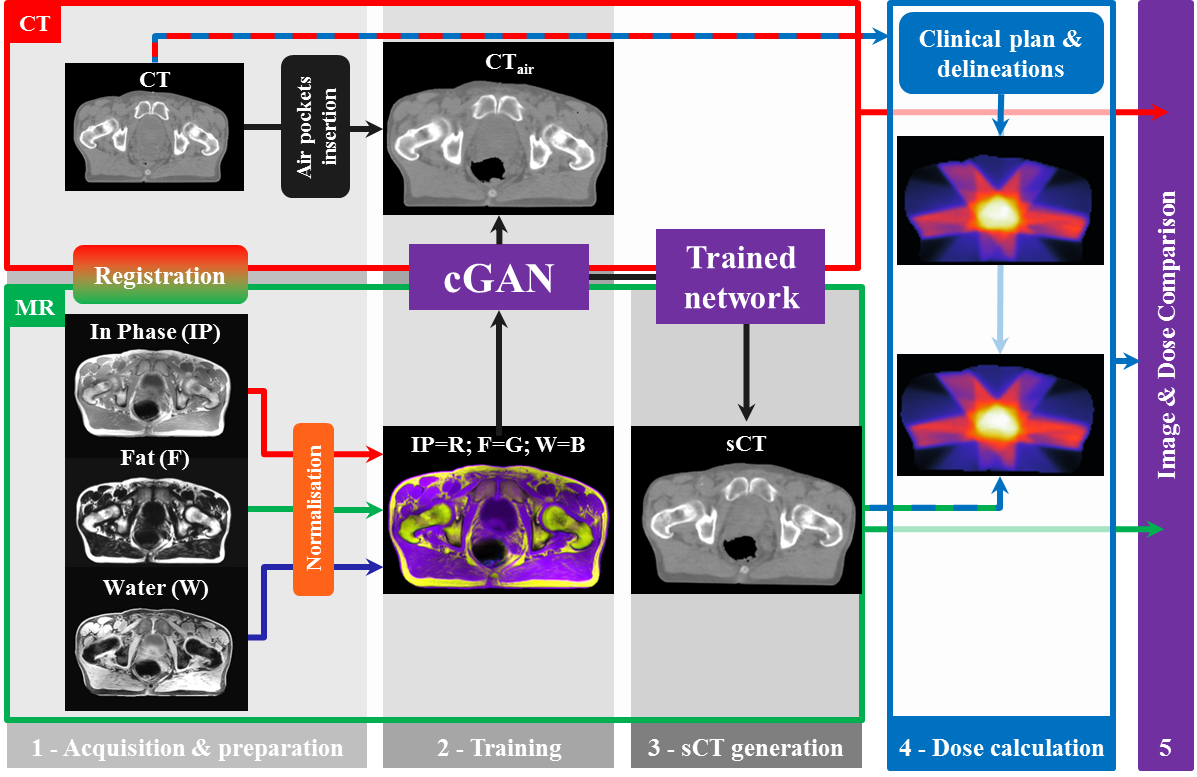}
\vspace{-5pt}
\caption[]{\label{fig:fig1} Schematic of the study. 1) After the acquisition of CT and MR images,
CT images were registered to MRI. CT and MR images were cropped in the inferior-superior direction to uniform the FOV. Image normalisation was also performed.
2) The 2D conditional generative adversarial network (cGAN) ``pix2pix'' was trained on 32 prostate patients and applied (feed-forward evaluation)
to the remaining 59 patients. For training and evaluation of the results, the location of air on MR was copied generating CT$_{\textrm{air}}$ to enforce
consistency in the location of air pockets.
The input of the cGAN was a coloured image using the three channels red, green and blue to accommodate in-phase, fat and water MR images, respectively.
3) The transverse plane was used for generation resulting in a volumetric sCT image after stacking the planes.
4) Dose planning and 5) image and dose evaluation were finally performed.}\vspace{-5pt}
\end{center}
\end{figure}

A cGAN called ``pix2pix'' consisting of a 256x256 U-net generator network and a 70x70 PatchGAN discriminator architecture was employed as
provided in the PyTorch implementation by \citeasnoun{Isola2016}.
As a proof-of-concept, considering that our main goal is the dosimetric evaluation of sCT images generated with a generative adversarial network,
we kept the network implementation as similar as possible to what was originally presented by Isola and co-workers.
Optimisation was performed as in \citeasnoun{Goodfellow2014} alternating between one gradient descendent step on the discriminator network
and one step on the generator network. A structured loss function cGAN$+\lambda \cdot$L1 with $\lambda$=100 was adopted.
As already investigated by \citeasnoun{Mathieu2015} and \citeasnoun{Isola2016}, the use of a loss function constituted by L1 alone leads to reasonable but blurred results; on the other hand,
cGAN alone will lead to sharp results but introducing artefacts in the images.
\citeasnoun{Isola2016} showed that training in an adversarial setting together with an L1 norm generates sharp images with a low amount of artefacts.
The weights of the network were randomly initialised from a Gaussian distribution with mean 0 and standard deviation 0.02.
The implementation of pix2pix can be applied to 8-bit grey-scale (1 channel) or coloured (3 channels) two-dimensional (2D) images.
All the patient data underwent pre-processing to normalise the MR images and prepare a paired experiment by registering input (MRI) and target data (CT).
In this work, we hypothesised that maximising the number of input images enriches contextual information per subject and facilitates learning of the image-to-image relation between MRI and CT.
Therefore, we used all the available multi-contrast
MR images to generate sCT images:
in-phase, fat and water images were coded as colours of the images (red, green and blue) and input of the network.
A schematic representation of the study is presented in Fig.~\ref{fig:fig1}.

\subsection{sCT generation}
\subsubsection{Image pre-processing} 
First, all the CT images were rigidly registered and resampled to MR with Elastix v4.7 \cite{Elastix}.
To keep the FOV consistent between CT and MR images, images were cropped in the superior-inferior direction to the smallest FOV ensuring that
training may be conducted in a paired fashion.
In the following, we use the term CT$_{\textrm{reg}}$, IP, F and W to refer the cropped images of registered CT, in-phase, water and fat images, respectively.
Before feeding the images to the cGAN, the voxel intensity of CT$_{\textrm{reg}}$ was clipped within the interval [$-$1000;1047]~HU
to avoid a too large discretisation step after conversion to 8-bits. 
Also, MR images were normalised to their 95\% intensity interval over the whole patient. Finally, all the images were converted to 8-bits to conform the pix2pix
implementation.

Before training, we tried to remove the impact of mismatch of air pockets location, e.g. in the rectum and bowel loops and enforce that air pockets were consistently located between CT$_{\textrm{reg}}$ and MRI. For this purpose,
air cavities were filled in CT$_{\textrm{reg}}$ and bulk-assigned ($-1000$~HU) as located in MR images using an automatic method
previously described by \citeasnoun{Maspero2017proton}. 
Without this pre-processing step, the generated sCT images may be characterised by inconsistent depiction of air between MR and sCT images, as reported in the supplementary material (LINK). 
Manual inspection was performed to verify correct assignment of the air on CT.
In the following, we use the term CT$_{\textrm{air}}$ to refer to these datasets. 
The generation of CT$_{\textrm{air}}$ has been introduced after noticing that the location of air when training directly on CT$_{\textrm{reg}}$ was not consistent with
MR images.

\subsubsection{Training of the network}
Training of the cGAN in a paired fashion was performed in the transverse plane randomly selecting 32 prostate cancer patients (training set).
The network was trained for 200 epochs on a Tesla P100 (NVIDIA, California, USA) graphical processing unit (GPU) with batch size of one.
Data augmentation was applied during training by flipping the images left and right and randomly cropping input and corresponding output images.

To verify the need for enforcing the consistency of the air location between CT and MRI during training, we repeated training using CT$_{\textrm{reg}}$ as a target of
the network. 

\subsubsection{Image generation}
The sCT generation was performed by applying the trained generator model and stacking all generated 2D transverse planes for each patient not used during training.
This patient group is considered as the ``test'' set. The 3D volumes were further post-processed using
DCMTK (\href{http://dicom.offis.de/dcmtk.php.en}{http://dicom.offis.de/dcmtk.php.en})
to create files that conform to the DICOM standard, usable in a treatment planning system.

Also, sCT generation for the prostate cancer patients in the test set (27 patients) was performed applying the generator trained using CT$_{\textrm{reg}}$ to evaluate the impact of
enforcing the consistency of the air location during training. In the following, we use the term sCT$_{\textrm{NoAir}}$ to refer to this dataset.

\subsection{Evaluation}
\label{subsec:Eval} 
The performance of the network was evaluated reporting the time needed to train the cGAN and to infer the generator on a GPU
and a central processing unit (CPU) framework (quad-core Intel Xeon 3.4~GHz).
\begin{figure}[!ht]
\begin{center}
\includegraphics[width=1.05\linewidth]{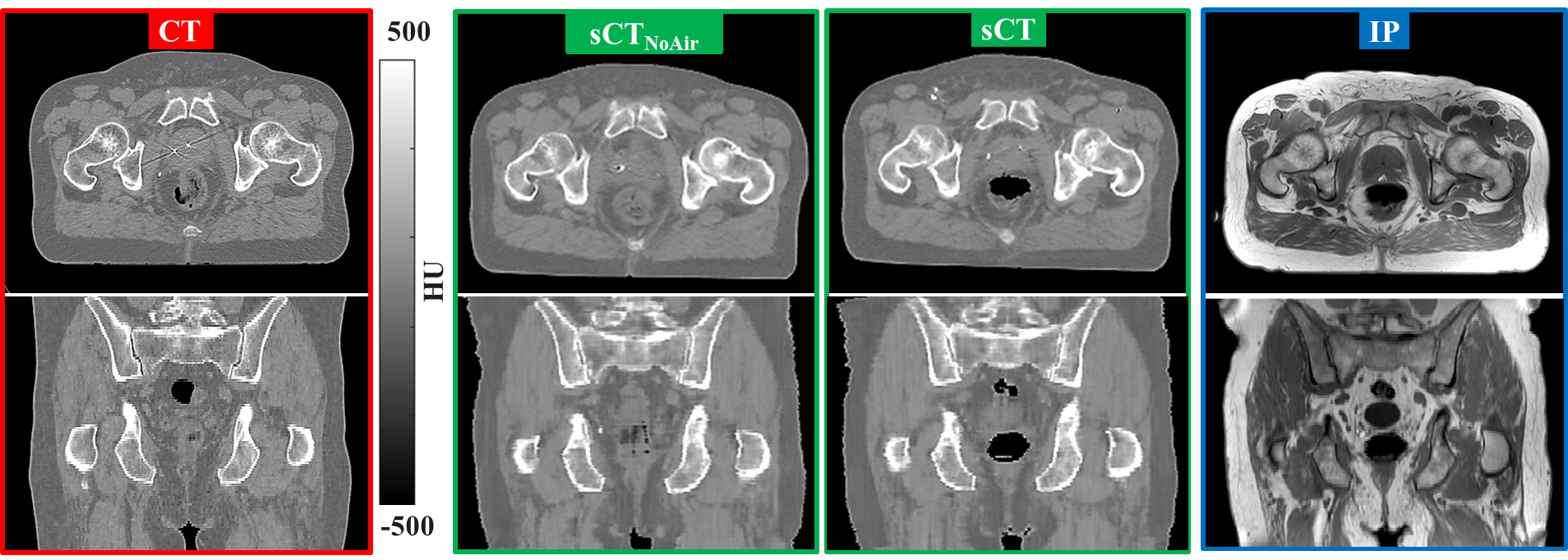}
\vspace{-15pt}
\caption[]{\label{fig:figCT_sCT} CT (left), sCT$_{\textrm{NoAir}}$ (middle, left) sCT (middle, right) and IP (right) for an exemplary patient treated for prostate cancer (top and central rows). The images on the top and bottom rows present a corresponding transverse plane, while in the bottom a corresponding coronal plane.}\vspace{-10pt}
\end{center}
\end{figure}

\subsubsection{Image comparison}

Image evaluation was performed in the test set by calculating the mean absolute error (MAE) and mean error (ME) on sCT with respect to CT$_{\textrm{reg}}$ (CT$_{\textrm{reg}}$ minus sCT) in
the intersection of the body contours. The body contours were automatically calculated by thresholding CT$_{\textrm{reg}}$ and sCT images at $-$500 HU.

In addition, MAE and ME were also calculated for the prostate patients in the test set (27) between sCT$_{\textrm{NoAir}}$ and  CT$_{\textrm{reg}}$.

\subsubsection{Dose comparison}
Thirty patients (ten for each tumour type) were randomly selected from the test set to undergo dose comparison.
For these patients, clinical plans were recalculated (QA modality) on sCT images in Monaco (v~5.11.02, Elekta AB, Sweden) using the Montecarlo photon algorithm on
a grid of 3x3x3~mm$^3$ with 3\% statistical uncertainty for VMAT plans (rectum and cervix) and 1\% for IMRT plans (prostate).

Before planning, the sCT images were rigidly registered, resampled and linearly interpolated to the planning CT using the inverse transformation of the 
registration that was previously found between planning CT and MR images.
The delineations used on the planning CT were also adopted for sCT except for the body contour, which was recalculated.

In almost all cervical cancer patients, except the two without integrated boost, the FOV acquired during MRI was insufficient to include all OARs and calculate the plan on
the patient. 
This was expected considering that MRI in the clinical settings is used for delineation of the sole primary tumour, while the plans may also include a boost to nodes in the common iliac and
para-aortic regions. We, therefore, performed dose recalculation for the two patients that had comparable FOV in the inferior-superior direction, and for the remaining eight patients
a new plan on the sCT was calculated prescribing 25x1.8~Gy to the PTV, excluding pathological nodes. The obtained plan was then recalculated on CT images.

Dose distributions were analysed through dose differences ($\frac{\textrm{CT}-\textrm{sCT}}{\textrm{prescribed dose}}$) and gamma analysis at
3\%,3mm and 2\%,2mm \cite{Low2010} within dose threshold regions of 90\%, 50\%, and 10\% prescription dose and in the body contour intersection
after a 15~mm cropping to exclude dose build-up.
Analysis of dose-volume histogram (DVH) points was performed to verify target (CTV, PTV) dose coverage and adherence to OARs
constraints considering the differences between dose points (D$_{98}$, D$_{50}$, D$_{2}$, V$_{95}$, V$_{75}$) on the CT and sCT plans for CTV, PTV and rectum for prostate patients
or bladder for rectal and cervical cancer patients. Note that DVH on sCT images were calculated on structures that were propagated after rigid registration; therefore, 
the DVH analysis does not take into account inter-scan differences of the structures.

\section{Results}
\label{sec:results}
\subsection{Performance of the network}
In total, 3495 transverse planes were used for training, which required about 11 hours on the GPU.
After cropping the FOV on CT and MR images, each patient was trained using a volume of about 109 transverse planes on average over the training set.
After training, inferring the generator network to obtain sCT images was performed in 5.6 s per patient on GPU and four times longer (21 s) on CPU.
Fig.~\ref{fig:figCT_sCT} presents the transverse (top) and coronal (center) planes of CT$_{\textrm{reg}}$ (left), sCT (middle) and IP (right)
for an exemplary prostate patient. It can be noticed that the different rectal filling between CT$_{\textrm{reg}}$ and IP is consistent to filling in the sCT image.
In the supplementary material, corresponding figures for exemplary rectal and cervical cancer patients are reported.

\begin{table}[!hb]
\renewcommand{\tabcolsep}{0.1cm}
\caption{Statistics of the image comparison between patients in the test set with prostate, rectal and cervical cancer in terms of mean absolute error (MAE) and mean error (ME)
between CT$_{\textrm{reg}}$ or CT$_{\textrm{air}}$ and sCT (and s$CT_{\textrm{NoAir}}$) over the intersection of the body contours.
The values are reported in terms of average ($\pm1\sigma$) and range [min; max] and expressed in Hounsfield Units [HU].}
\begin{center}
       \small{
      \begin{tabular}{c cc cc }
        \toprule
     \textbf{Tumour}  &\textbf{Number} &  \multirow{2}{*}{\textbf{Data}} & MAE & ME \\ 
        \textbf{location}  &\textbf{of patients} &  & [HU] & [HU] \\  
        \midrule
        &  &  \multirow{2}{*}{CT$_{\textrm{reg}}-$sCT} & 65$\pm$10 & \ \ 1$\pm$6  \\
&  & & [50;97] & [$-$12;15]    \\
    & &\crule{3}\\
  \multirow{2}{*}{\textbf{Prostate}}   & \multirow{2}{*}{27} &  \multirow{2}{*}{CT$_{\textrm{air}}-$sCT} & 60$\pm$6 \ & \ $-$3$\pm$5  \\
&  & & [48;71] & [$-$18;8]  \\  
& & \crule{3}\\
  & &  \multirow{2}{*}{\ \ \ \ \ \ \ CT$_{\textrm{reg}}-$sCT$_{\textrm{NoAir}}$\ } & 65$\pm$9 \ & \ $-$2$\pm$10 \\
&  & & [51;95] & [$-$21;13]    \\  
       \midrule              
\multirow{2}{*}{\textbf{Rectum}}  &  \multirow{ 2}{*}{18} &  \multirow{2}{*}{CT$_{\textrm{reg}}-$sCT}& 56$\pm$5 \ & \ \ 2$\pm$9  \\
& & &  [48;67] & [$-$16;23]    \\
        \midrule              
\multirow{ 2}{*}{\textbf{Cervix}}  & \multirow{ 2}{*}{14} &  \multirow{2}{*}{CT$_{\textrm{reg}}-$sCT} & 59$\pm$6 \ & \ \ \ 4$\pm$10  \\
& & &  [50;69] & [$-$16;22]    \\
\bottomrule
\end{tabular}
}
\end{center}
\label{tbl:ImageDiff}
\end{table}
\subsection{Image comparison}
Tab.~\ref{tbl:ImageDiff} reports the statistics of image comparison in terms of MAE and ME in the patient test
set between CT$_{\textrm{reg}}$/CT$_{\textrm{air}}$ and sCT.
Over the entire test set (59), the MAE and ME were, on average, 61$\pm$9 and 2$\pm$8 HU.
It can be noticed that the MAE and ME are comparable among patients with different tumour location and the MAE decreases when comparing sCT to
CT$_{\textrm{air}}$. This demonstrates that CT$_{\textrm{air}}$ is more similar to sCT than CT$_{\textrm{reg}}$, which justifies its
use during the training of the network in a paired fashion.

When considering MEA and ME of the network trained with and withour enforcing air consistency, we observer that the metrics  are comparable in the two scenarios, but the ME slightly decreases without enforced air location consistency.
This result may be explained by the fact that the size of air pockets is much smaller than the entire body contour. In this sense,
voxelwise differences to the air location are not expected to greatly impact the reported MAE and ME. Nevertheless, if we consider Fig.~\ref{fig:figCT_sCT}, 
we can observe that air is filled with soft tissue in case training was performed without enforcing air consistency.

\begin{table}[!hb]
\renewcommand{\tabcolsep}{0.1cm}
\caption{Statistics of the 10/27 prostate, 10/18 rectal and 10/14 cervical cancer patients among the test set.
Mean dose difference relative to the prescribed dose and gamma pass rate among the average dose difference calculated on
a threshold of 10\%, 50\%, and 90\% of the prescribed dose 
and the intersection of the body contour between CT and sCT images (Body). The values are reported in terms of average ($\pm1\sigma$) and range [min; max].}
\begin{center}
       \small{
      \begin{tabular}{c c ccc }
        \toprule
     \textbf{Tumour}  &\textbf{Volume} & Dose Difference & Pass Rate & Pass Rate \\ 
        \textbf{location}  &\textbf{of interest} & $\frac{\mathrm{CT}-\mathrm{sCT}}{\mathrm{D}_{\mathrm{Prescr}}}$ & $\gamma_{3\%,3\,\mathrm{mm}}$ & $\gamma_{2\%,2\,\mathrm{mm}}$ \\       
      & & [\%] & [\%] & [\%] \\ 
      \midrule
& D$>10\%$ &  $-0.1\pm0.1$ & $98.1\pm1.2$ & $95.0\pm2.3$   \\
       & & $[-0.3;0.2]$ & \ $[96.3;99.4]$ & $[91.4;97.8]$  \\
  & D$>50\%$&  $-0.1\pm0.2$ & $99.4\pm0.6$ & $97.4\pm1.6$   \\
\textbf{Prostate}        & & $[-0.4;0.5]$ & $[98.1;100]$ & $[93.8;99.7]$  \\
(10 patients) & D$>90\%$ &  $-0.3\pm0.4$ & $99.7\pm0.2$ & $97.6\pm2.3$   \\
       & & $[-1.1;0.4]$ & $[99.3;100]$ & $[91.8;99.9]$  \\ 
       & Body &  $0.0\pm0.1$ & $98.8\pm0.7$ & $96.8\pm2.1$   \\
       & & $[-0.2;0.1]$ & $[97.5;99.6]$ & $[93.2;98.9]$  \\
       \midrule              
                  & D$>10\%$ &  $-0.2\pm0.2$ & $97.1\pm1.1$ & $91.6\pm3.3$   \\
   & D$>50\%$&  $-0.3\pm0.3$ & $98.5\pm1.1$ & $93.2\pm3.6$   \\
\textbf{Rectum}  & & $[-0.8;0.0]$ & $[96.0;99.7]$ & $[86.5;97.7]$  \\
(10 patients)    & D$>90\%$ &  $-0.3\pm0.5$ & $98.5\pm2.1$ & $92.0\pm6.6$   \\
       & & $[-1.0;0.6]$ & $[93.2;99.9]$ & $[77.9;98.2]$  \\ 
       & Body &  $-0.2\pm0.1$ & $97.6\pm1.2$ & $94.0\pm3.0$   \\
       & & $[-0.4;-0.1]$ & $[95.7;98.8]$ & $[88.7;96.8]$  \\
       \midrule              
                  & D$>10\%$ &  $-0.1\pm0.3$ & $97.1\pm1.7$ & $92.9\pm3.7$   \\
       & & $[-0.6;0.2]$ & $[93.8;98.7]$ & $[84.3;96.1]$  \\
   & D$>50\%$&  $-0.2\pm0.5$ & $99.6\pm1.9$ & $94.5\pm4.6$   \\
\textbf{Cervix}       & & $[-1.5;0.4]$ & $[94.0;100]$ & $[83.1;98.6]$  \\
(10 patients)    & D$>90\%$ &  $-0.1\pm0.7$ & $98.5\pm3.1$ & $90.6\pm6.8$   \\
       & & $[-1.6;1.0]$ & $[89.9;99.9]$ & $[72.3;96.3]$  \\ 
       & Body &  $-0.1\pm0.3$ & $97.7\pm1.7$ & $93.6\pm4.0$   \\
       & & $[-0.9;0.2]$ & $[94.3;99.4]$ & $[86.6;98.1]$  \\        
\bottomrule
\end{tabular}
}
\end{center}
\label{tbl:DoseDiff}
\end{table}

\subsection{Dose comparison}
\begin{figure}[!ht]
\begin{center}
\includegraphics[width=1\linewidth]{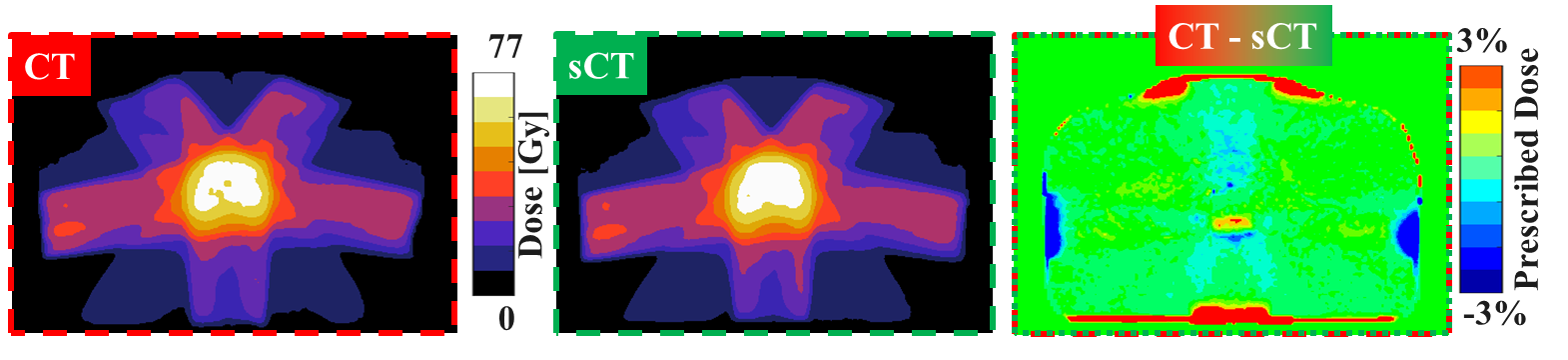}
\vspace{-15pt}
\caption[]{\label{fig:figDose_CT_sCT} Doses calculated on CT$_{\textrm{reg}}$ (left)
and sCT (middle) in the transverse plane corresponding to the isocentre for one prostate cancer patient. (Right) The dose difference (CT$_{\textrm{reg}}$-sCT) is presented as the percentage of the prescribed dose (77 Gy) for the corresponding plane.}\vspace{-10pt}
\end{center}
\end{figure}

An example of dose calculated on CT$_{\textrm{reg}}$ and sCT along with their difference
is presented in Fig.~\ref{fig:figDose_CT_sCT} for the same prostate patient as shown in Fig.~\ref{fig:figCT_sCT}.
On average (Tab.~\ref{tbl:DoseDiff}), it was observed that sCT images result in a higher dose to the target of about 0.1-0.3\%.
In the worst case, the mean dose difference was 1.6\%.
The average gamma pass rates using the 3\%,3mm and 2\%,2mm criteria were $>$ 97 and 91\%, respectively, for all volumes of interests considered.

As part of the supplementary material, the dose difference of each individual patient in the high dose region (D$>90\%$) is presented
for all thirty patients included in the dose comparison.

\begin{figure}[!t]
\begin{center}
\begin{subfigure}{}
\includegraphics[width=\linewidth]{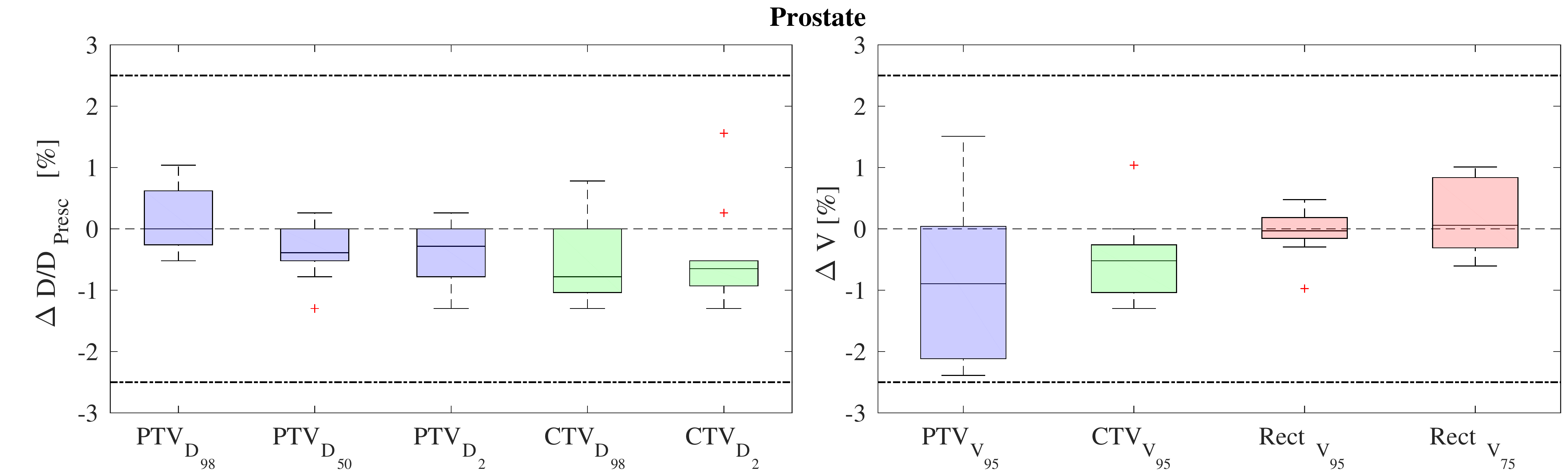}
\end{subfigure}
\vspace{-20pt}
\begin{subfigure}{}
\includegraphics[width=\linewidth]{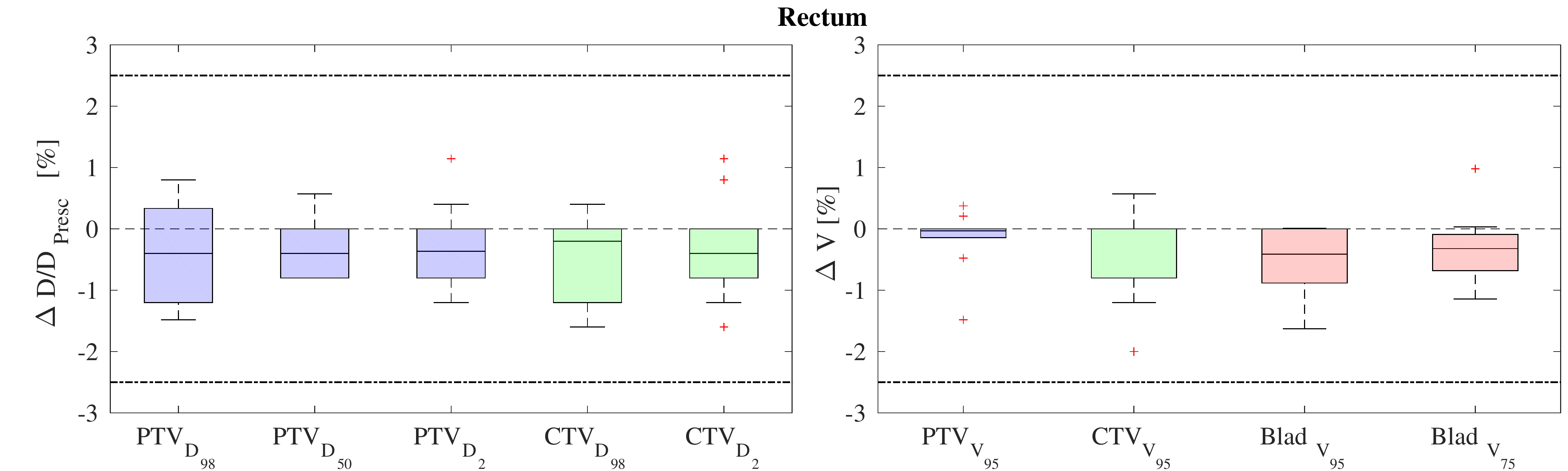}
\end{subfigure}
   \vspace{-20pt}
  \begin{subfigure}{}
  \includegraphics[width=\linewidth]{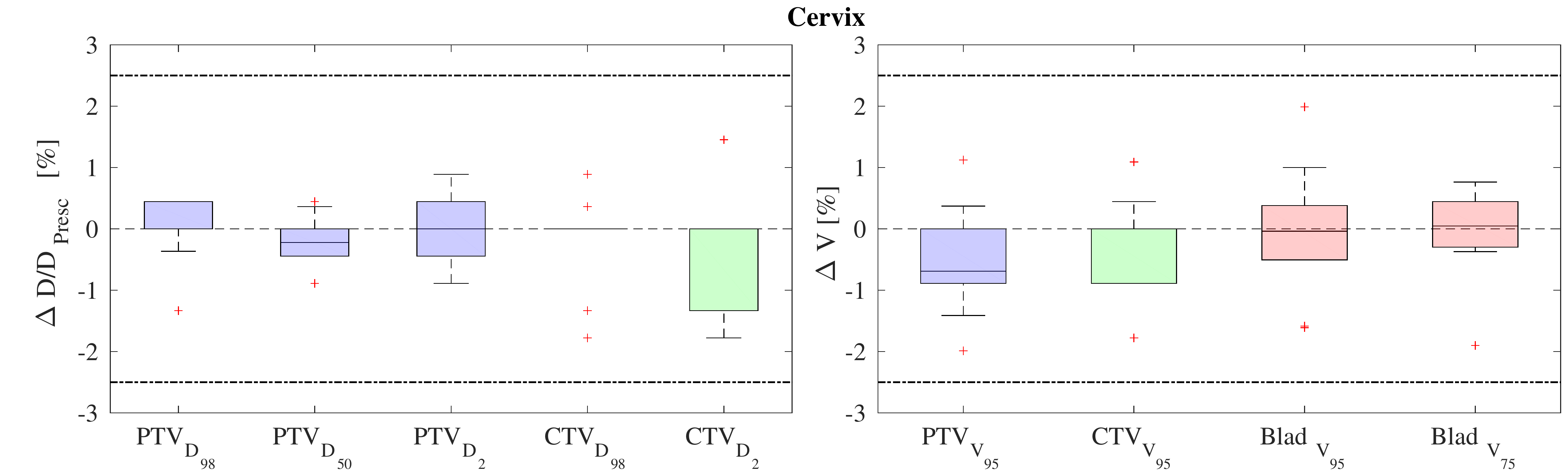}
\end{subfigure}
\vspace{-15pt}
\caption[]{\label{fig:fig3} Boxplots of targets (CTV, PTV) and OARs DVH parameter differences between dose on CT and sCT (CT$-$sCT) for the prostate (top),
rectal (middle) and cervical (bottom) cancer patients.
The values refer to the whole course of fractionated treatment and are rescaled to the prescribed dose (left) or the total volume of the specific structure (right).}
\vspace{-5pt}
\end{center}
\end{figure}
Fig.~\ref{fig:fig3} presents the boxplot of the DVH point difference (CT$-$sCT) for targets (PTV, CTV) and OARs showing that all the DVH points on sCT-derived plans
were within $\pm$2.5\% with respect to the corresponding points on CT-derived plans. 

\section{Discussion}

Here, we demonstrated that deep learning enabled fast generation of sCT images facilitating accurate MR-based dose calculation for irradiation of patients with cancer in the pelvic area. 
In particular, we showed that training a cGAN on prostate cancer patients results in MR-based dose calculation within  an average dose difference of 0.5\% compared to 
CT-based calculations (CT$-$sCT). Though the network was trained on prostate cancer patients, sCT images generation
in rectal and cervical cancer patients also resulted in accurate dose calculations (see also the supplementary material).
This is of particular interest, considering that the network seems to accurately solve image-to-image translation problems also for 
female patients that were not included in the training set.
However, for the cervical patients, the FOV coverage is insufficient to recalculate clinical plans. This means that for this patient group the
MR sequence should be revised (i.e. extension of the FOV) and newly evaluated before clinical use.
Also, by enforcing consistency of air location during training, we demonstrated that the
network was able to depict internal air in location consistent to MRI.
Please note that a ``correct'' depiction of internal air is not crucial in the scope of standard radiotherapy, given 
its limited size and the fact that it may be differently located at each fraction. However, it may become of larger interests when considering 
MRgRT, given the additional dose at tissue interfaces \cite{Raaijmakers2008} and considering the scenario of hypofractionated treatments \cite{Benjamin2017}.

The dosimetric evaluation performed in this work has been restricted to the pelvic area.
The sCT generation method here adopted, however, could be applied also to different anatomical locations after proper training of the network.
This makes the approach generic.

Within this work, for the first time, sCT generated with a deep learning technique underwent 
dosimetric evaluation in the pelvic region. In general, the dose differences obtained in this study are in line with previously published studies on prostate cancer patients, which reported
dose differences within 1\% \cite{Dowling2015,Korhonen2014,Kim2015,Siversson2015,Prior2016,Tyagi2016,Maspero2017Quant,Persson2017}, and to other studies
in the pelvic area \cite{Kemppainen2017,Liu2017,Wang2018}. 
Dose deviations should be interpreted in the context of the clinically acceptable uncertainty in radiation therapy. When considering the complete radiotherapy pathway,
including uncertainties in beam calibration, relative
dosimetry, dose calculations, and dose delivery, the International Commission on
Radiation Protection estimated an uncertainty of 5\% in a clinical set-up \cite{ICRP2000,Thwaites2013}.
The dosimetric deviation of an MR-based dose calculation (assuming CT to be the ground truth)
of less than 0.5\% only makes up for a small fraction of the total uncertainty \cite{Persson2017}. 

In this study, a conventional multi-gradient echo sequence was employed as already used by \citeasnoun{Tyagi2016}, \citeasnoun{Maspero2017Quant}, \citeasnoun{Kemppainen2017}.
This is the first time that multi-contrast MR images acquired with a single sequence have been used as input of a deep learning network for sCT generation.
Previous work showed that multi-contrast images from different sequences, e.g. Dixon and ultra-short echo time could be used for generating images for 
attenuation correction for MR-PET \cite{Leynes2018}.
It is still unclear whether the use of multi-contrast images effectively increase the quality sCT generation or facilitate the training of the network.
Future investigations may clarify this aspect, however, based on our findings,
we believe that the use of deep learning makes specialised MR sequences, e.g. ultra-short echo time for direct bone visualisation, obsolete.
This may lower the requirements for MR sequences used for sCT generation as well as the acquisition time.
Of course, a high geometric fidelity, e.g. by means of high bandwidth, of the sequence is still required as adopted in this work.
In particular, for the sCT generation method here proposed, evaluation of geometric accuracy, especially in the case these sCT images may be used for position verification
purposes is still required.

In general, lowering requirements for the quality of MR sequence used for sCT generation may be of particular interest for MRgRT. 
For example, investigating the use of accelerated MR imaging techniques \cite{Feng2014,Brix2014,Hollingsworth2015} to speed up the acquisition time may facilitate an MR-based dose calculation also for locations affected by high tissue mobility, which is a currently unmet need of MR-only radiotherapy \cite{Owrangi2018}.

In this study, we enforced consistency of patient anatomy during training by performing rigid registrations and assigning air locations from MRI to CT.
As an alternative approach, non-rigid registrations could have been adopted.
However, we decided to avoid this approach since it would
have introduced additional geometrical uncertainties, due to registration errors \cite{Thor2011,Thor2013}, that we preferred not to introduce in this study.
Also, non-rigid registration could have masked possible image deformation that are inherent in MR images \cite{Fransson2001,Wang2004,Walker2014}.
Future studies are advocated to clarify this aspect.
Also, it is of interest to investigate the use of GANs in an unpaired fashion \cite{Zhu2017} since it may eliminate the need of a perfectly aligned dataset in the
training phase as shown by \citeasnoun{Wolterink2017} for brain cancer patients. 
Training in an unpaired fashion may be of particular interest since it may avoid the need of minimising
inter-scan differences by copying air pockets from sCT to CT. Moreover, a 2.5D or 3D network could also be investigated to solve the discontinuities observed in
the inferior-superior direction after stacking the transverse planes.

The sCT generation was performed in less than 6~s for a single patient volume. 
The computation time can be affected by many factors, e.g. the type of GPU adopted, the matrix and FOV size. When compared to existing methods, the sCT images presented in our work are 
generated faster even when compared to other deep-learning based methods \cite{Han2017,Wolterink2017,Dinkla2018}
This can facilitate daily sCT generation for application where time constraints are crucial, e.g. in MRgRT \cite{Lagendijk2014}.
A limitation of the current study is that MR-based dose calculations were assessed in the absence of magnetic field. For MRgRT, dose calculations require particular attention due to
the presence of a magnetic field affecting the dose distribution, especially near air cavities \cite{Raaijmakers2008}.
Given the promising results, a future study will investigate whether MR-based dose calculation on sCT obtained from the same network can be considered accurate
also in the presence of magnetic fields.
Also, before clinical usage in a complete MR-only workflow, the sCT generation method still needs to be thoroughly tested for accuracy in position verification.
Moreover, a safe clinical implementation may also require designing quality assurance methods to validate the sCT images in the absence of the ``gold standard''
offered by CT.

\section{Conclusion}
\label{sec:Concl}
To conclude, this study shows, for the first time, that sCT images generated with a deep learning approach employing a cGAN
and multi-contrast MR images acquired with a single acquisition facilitated accurate dose calculations in prostate cancer patients.
It was further shown that without retraining the network, the cGAN could generate sCT images in the pelvic region for accurate dose calculations for rectal and cervical cancer patients.
A particularly attractive feature of our method is its speed as it allows sCT generation within 6 seconds on a GPU and within 21 seconds on a CPU.
This could be of particular benefit for MRgRT applications.

\section*{Acknowledgements}
\addcontentsline{toc}{section}{Acknowledgements}
Max A Viergever and Jan J W Lagendijk offered general support to the research.\\
The authors would like to thank Gert J Meijer, Hans C J de Boer, Filipa Guerreiro, Daan M de Muinck Keizer and Christopher Kurz
for the inspiring discussions in the MR-only group meeting at the University Medical Center Utrecht.

The research is funded by ZonMw IMDI Programme (project number: 1040030).
The study was classified as non-WMO by the medical ethical commission (Medisch Ethische Toetsingscommissie) under the protocol number 15-444/C.\\
Peter R Seevinck declares to be a minority shareholder of MRIguidance BV.\\
Cornelis A T van den Berg declares to be a minority shareholder of MRCode BV.


\section*{References}
\addcontentsline{toc}{section}{References}
\bibliography{Biblio_cGaN}

\end{document}